\documentstyle[amssymb,aps,multicol,epsfig,fancyheadings]{revtex}

\begin{document}
\title{Submillimeter spectroscopy of tilted Nd$_{1.85}$Ce$_{0.15}$CuO$_{4-\delta }$
films: observation of a mixed a-c excitation}
\author{A. Pimenov$^{1}$, A. V. Pronin$^{1}$, A. Loidl$^{1}$, A. Kampf$^{2}$, S. I.
Krasnosvobodtsev$^{3}$, and V. S. Nozdrin$^{3}$}
\address{$^{1}$Experimentalphysik V, EKM, Universit\"{a}t Augsburg, 86135 Augsburg,
Germany\\
$^{2}$Theoretische Physik III, EKM, Universit\"{a}t Augsburg, 86135
Augsburg, Germany\\
$^{3}$Lebedev Physics Institute, Russ. Acad. Sciences, 117942 Moscow, Russia%
}
\maketitle

\begin{abstract}
The anisotropic conductivity of a series of tilted Nd$_{1.85}$Ce$_{0.15}$CuO$%
_{4-\delta}$ thin films was measured by quasioptical spectroscopy in the
frequency range 3\,cm$^{-1}<\nu <40$\,cm$^{-1}$. Two characteristic features
have been observed in the low-temperature transmission spectra. The first
one at $\nu = 12$\,cm$^{-1}$ was shown to reflect the c-axis plasma
frequency of Nd$_{1.85}$Ce$_{0.15}$CuO$_{4-\delta}$. The second feature
represents a new mixed ab-plane/c-axis excitation. The frequency of this
resonance may be changed in a controllable way by rotating the polarization
of the incident radiation.
\end{abstract}


\begin{multicols}{2}

Among the cuprate superconductors, Nd$_{2-x}$Ce$_{x}$CuO$_{4}$ (NCCO) has
attracted considerable interest because it can be considered as
electron-doped compound\cite{tokura}, which reveals the (T$^{\prime }$)
tetragonal structure and is extremely two-dimensional\cite{tokura,hidaka}.
Earlier in-plane microwave experiments on NCCO \cite{andreone,anlage,wu}
appeared to be consistent with the conventional s-wave BCS predictions.
However, recent data including flux quantization \cite{tsuei} and
penetration depth at microwaves \cite{kokales} and at radiowaves \cite
{prozorov} provided strong experimental evidence of d-wave symmetry of the
pairing state.

In contrast to a number of ab-plane experiments, there exists only few
information concerning the c-axis properties of NCCO due to small dimensions
of the samples along the c-axis. Most experiments on c-axis dynamics\cite
{shibata,heyen} were carried out using ceramic NCCO samples. Alternatively,
the method of grazing-incidence reflection is known \ to be powerful in
extracting the c-axis properties of strongly anisotropic superconductors\cite
{marel}. The experiments using instead of the grazing reflection a tilted
sample\cite{markowitsch} and the normal-incidence geometry may be considered
as modification of this method.

In this letter we present the results of submillimeter-wave (3\,cm$^{-1}<\nu
<40$\,cm$^{-1}$) experiments on tilted Nd$_{1.85}$Ce$_{0.15}$CuO$_{4-\delta}$
films at low temperatures. The spectra reveal two new features below the
superconducting transition, which could be identified as: i) plasma
resonance along the c-axis and ii) mixed ab-plane/c-axis excitation. Using
the measurements at different polarizations of the incident radiation it was
possible to separate the effective conductivity at a tilted angle into
components within the ab-plane and along the c-axis.

The experiments were carried out on different Nd$_{1.85}$Ce$_{0.15}$CuO$%
_{4-\delta }$ films yielding similar results. For simplicity, the results
from only one film are presented. The films were prepared using a
two-beam-laser deposition on yttrium stabilized ZrO$_{2}$ substrates\cite
{preparation}. X-ray analysis revealed the c-axis orientation of the films.
The substrate of the presented film was tilted from (001) plane by an angle $%
\alpha =0.8^{o}\pm 0.4^{o}$. Therefore, the film was also tilted by the same
angle from the ideal c-axis orientation. The ac-susceptibility measurements
revealed a narrow superconducting transition ($\Delta T[10\%-90\%]=0.9$%
\thinspace K) with the onset temperature of 20.1\thinspace K. The film
thickness, estimated from the deposition time, was $d=170$\thinspace nm.

The transmission experiments in the frequency range 6\thinspace cm$^{-1}<\nu
<40$\thinspace cm$^{-1}$ were carried out in a quasioptical arrangement \cite
{volkov}, which allows both the measurements of transmission and phase shift
of a film on a substrate. The properties of the substrate were determined in
a separate experiment. Utilizing the Fresnel optical formulas for the
complex transmission coefficient of the substrate-film system, the absolute
values of the complex conductivity $\sigma ^{*}=\sigma _{1}+i\sigma _{2}$
were determined directly from the observed spectra without any
approximations. The radiation in this method is continuously tunable in
frequency and is linearly polarized. This allows to perform experiments for
different directions of the currents, induced in the sample. The geometry of
the transmission experiment is represented in the inset of Fig. 1.

Fig.\,1 shows the transmission spectra of a Nd$_{1.85}$Ce$_{0.15}$CuO$%
_{4-\delta }$ thin film at $T=6$ K and at $T\approx T_{C}$. A new feature,
that can be observed in the spectra of the superconducting state, is the
appearance of a transmission peak at $\nu =12$\,cm$^{-1}$. This feature may
be called ''antiresonance'' as it corresponds to an {\em increase} in
transmission. In addition to this antiresonance, a suppression of the
low-temperature transmission is seen in the frequency range 18\,cm$^{-1}<\nu
<30$\,cm$^{-1}$. This suppression corresponds to a broad resonance and is
somewhat masked by the interference pattern, caused by the substrate.

Fig. 2 shows the effective conductivity of the film as calculated from
transmission and phase shift spectra for $T=$\thinspace 6\thinspace K.
Because the exact expressions for the complex transmission coefficient were
used, the substrate interference patterns are absent in this presentation.
The effective conductivity is shown for different orientations of incident
radiation and is labeled by the angle $\varphi $ (inset of Fig. 1).
\begin{figure}[th]
\centering
    \includegraphics[width=7cm,clip]{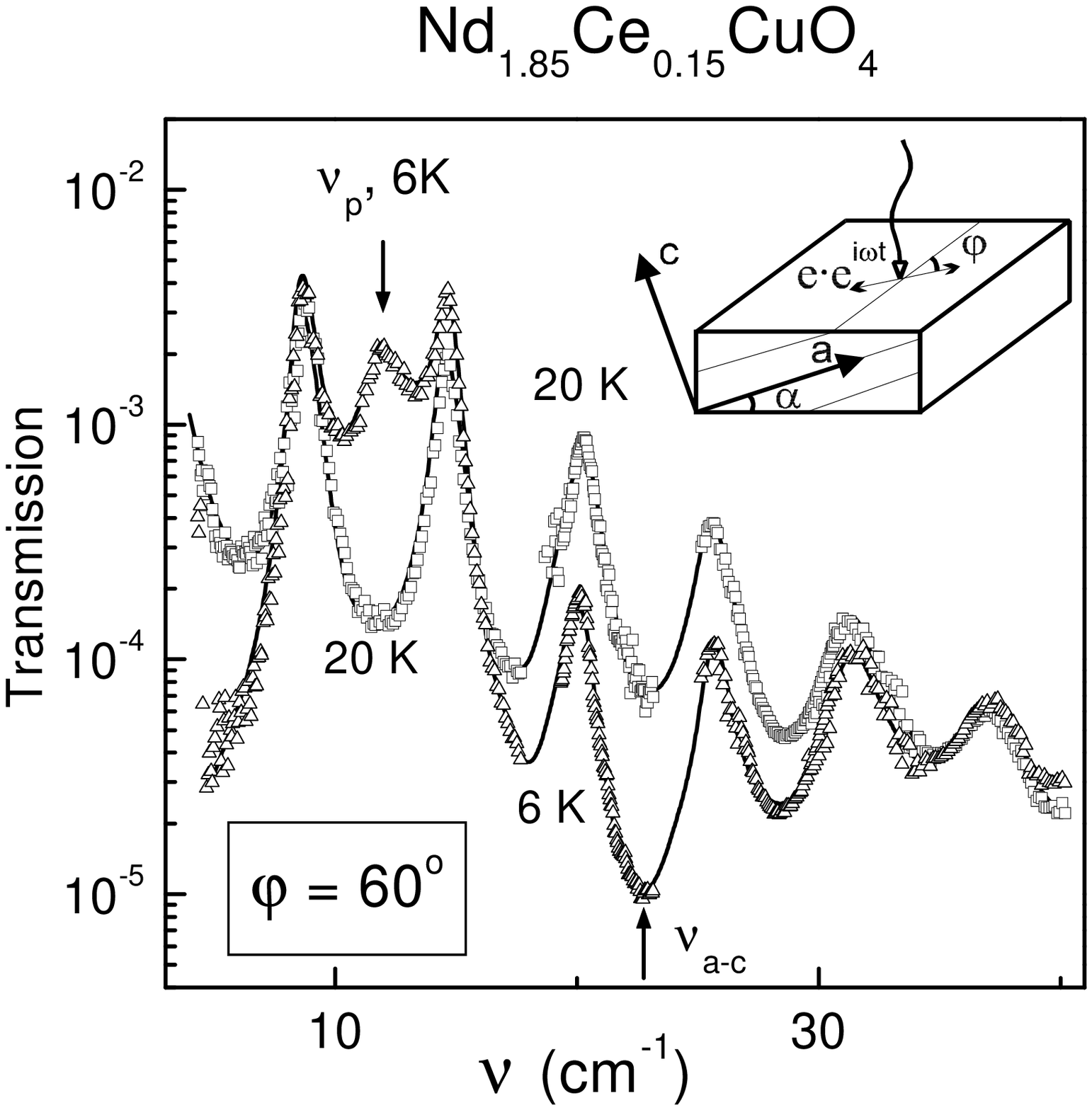}
    \vspace*{1mm}\caption{Submillimeter-wave transmission spectra of a Nd$_{1.85}$Ce$_{0.15}$%
CuO$_{4-\delta}$ film at $T=T_C$ (20K) and at T=6K. Lines are drawn to guide
the eye. Arrows indicate the position of the c-axis plasma frequency and of
a mixed a-c excitation. The inset shows the geometry of the transmission
experiment.}\label{figtr}
\end{figure}
\begin{figure}[th]
\centering
    \includegraphics[width=7cm,clip]{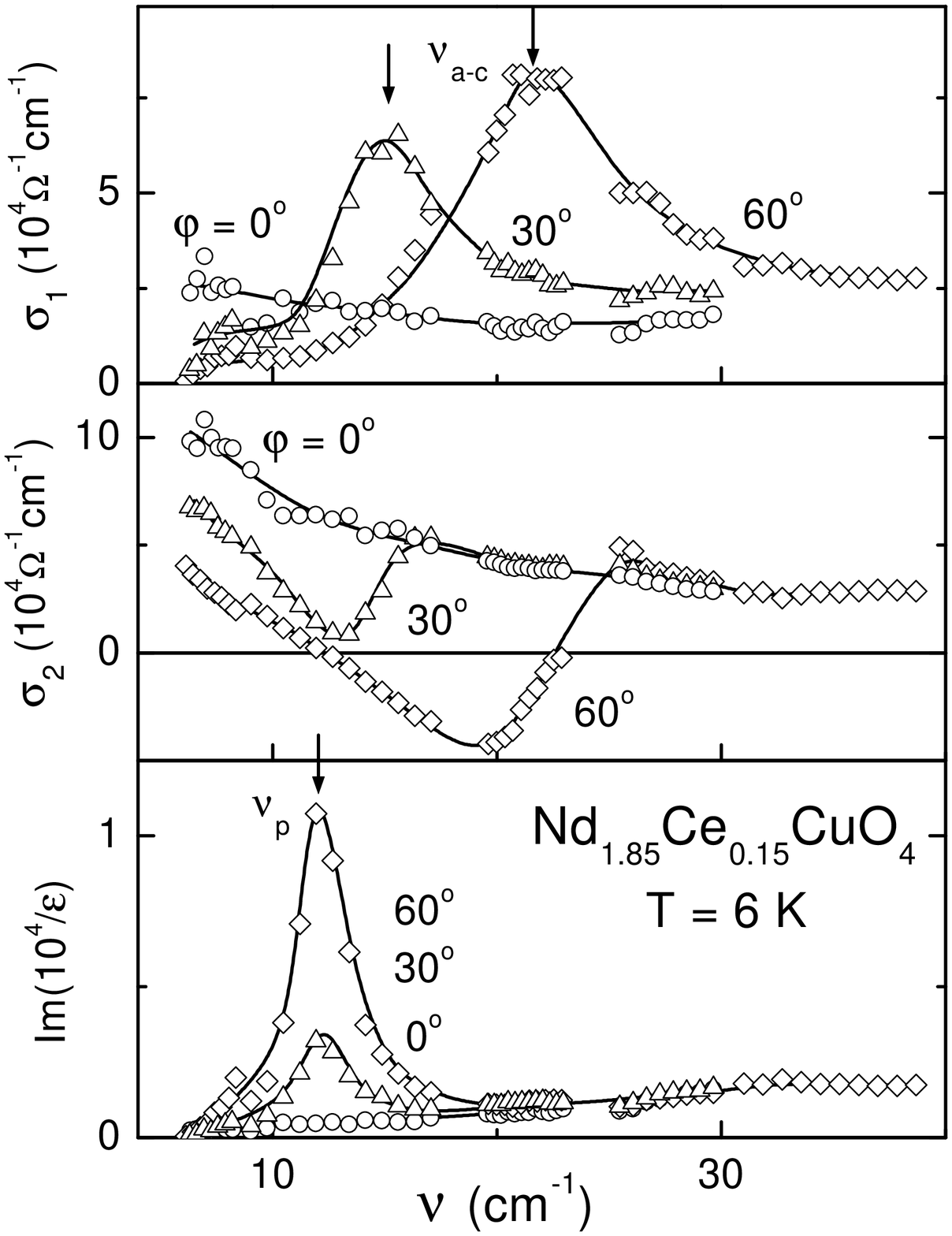}
    \vspace*{1mm}\caption{ Low-temperature conductivity spectra of a Nd$_{1.85}$Ce$_{0.15}$CuO%
$_{4-\delta}$ film for different orientations of the incident radiation.
Lines are drawn to guide the eye. Upper panel: $\sigma _1$. Middle panel: $%
\sigma _2$. Lower panel: Imaginary part of the loss function $%
Im(1/\varepsilon)$. Arrows indicate the position of the c-axis plasma
frequency and of the mixed a-c excitation.}
\label{figangle}
\end{figure}
The comparison of Fig. 1 and Fig. 2 clearly shows that a zero-crossing of $%
\sigma _{2}$ and a peak in the loss function Im$(1/\varepsilon )$ correspond
to the antiresonance at $\nu =12$\,cm$^{-1}$. The suppression of the
transmission at $\nu \sim 23$\,cm$^{-1}$ yields a peak in the real part of
conductivity. Interestingly, both features have completely different
dependence upon the rotating of the polarization plane ($\varphi $). As will
be seen below, the antiresonance at 12\,cm$^{-1}$ corresponds to the c-axis
plasma frequency and is therefore independent upon the polarization of the
incident radiation. Contrary, the peak at 23\,cm$^{-1}$ ($\varphi =60^{o}$)
represents the mixed ab-plane/c-axis excitation. The degree of mixing may be
controlled by the geometry of the experiment, which explains the $\varphi $%
-dependence of the peak position.

The spectra for $\varphi =0^{o}$ corresponds to currents flowing only within
the ab-planes and resemble the spectra of a superconductor below the gap
frequency: $\sigma _{1}$ has only a weak frequency dependence and is
suppressed by approximately a factor of three compared to the normal-state
conductivity, $\sigma _{2}$ is strongly enhanced compared to the normal
state and reveals approximately a $1/\omega $ frequency dependence. From its
slope the low-frequency penetration depth may be estimated as $\lambda
_{ab}(6K)=230\pm 30$ nm.

In order to calculate the effective conductivity of a tilted sample,
consider a free-standing film of thickness $d$ in a uniform electromagnetic
field $Ee^{-i\omega t}$ as shown in the inset of Fig. 1. For simplicity, the
film is assumed to be thin compared to the penetration depth, $d\ll \lambda $%
. In this case the current and field distribution may be considered to be
uniform. Taking into account the charges formed at the surface, the
following equation for the effective conductivity is obtained:

\begin{equation}
\sigma _{eff}=\frac{-i\varepsilon _{0}\omega (\sigma _{a}\cos ^{2}\alpha
+\sigma _{c}\sin ^{2}\alpha )+\sigma _{a}\sigma _{c}}{-i\varepsilon
_{0}\omega +\sigma _{a}\sin ^{2}\alpha +\sigma _{c}\cos ^{2}\alpha }
\label{eqsig}
\end{equation}
Here $\varepsilon _{0}$ is the permittivity of free space, $\sigma _{a}$ $%
(\sigma _{c})$ is the complex conductivity in the ab-plane (along the
c-axis), and $\omega$ is the angular frequency.

Within the approximation $\alpha \approx \sin \alpha \ll 1$ and $|\sigma
_{a}|\gg |\sigma _{c}|$, the Eq. (\ref{eqsig}) can be written as:

\begin{equation}
\sigma _{eff}=\frac{\sigma _{a}(\sigma _{c}-i\varepsilon _{0}\omega )}{%
\sigma _{a}\alpha ^{2}+(\sigma _{c}-i\varepsilon _{0}\omega )}
\label{eqsimpl}
\end{equation}

Two simple conclusions can be immediately drawn from this expression, if the
real parts of the conductivities $(\sigma _{1a},\sigma _{1c})$ are only
weakly frequency dependent: i) the effective conductivity should reveal a
peak if $Im[\sigma _{a}\alpha ^{2}+(\sigma _{c}-i\varepsilon _{0}\omega
_{1})]=0$ and ii) the inverse conductivity shows a peak (i.e. longitudinal
resonance) if $Im[\sigma _{c}-i\varepsilon _{0}\omega _{0}]=0$. Assuming
that only the high-frequency dielectric constant $(-i\varepsilon
_{0}\varepsilon _{c,\infty }\omega )$ and the superconducting condensate $%
(in_{s}e^{2}/m_{c}\omega )$ contribute to the low-temperature c-axis
conductivity, the critical frequency in the case ii) can be estimated $%
\omega _{0}^{2}=n_{s}e^{2}/[m_{c}(\varepsilon _{\infty }+1)]$, which closely
resembles the expression for the screened plasma frequency. Evidently, the
expression for $\omega _{0}^{2}$ does not dependent upon the tilt angle. On
the contrary, a substantial angular dependence is expected for the resonance
at $\omega _{1}$. The change of the tilt angle may be easily achieved
experimentally by rotating the polarization of the incident radiation. In
this case the effective tilt angle may be obtained as $\alpha _{eff}=\alpha
_{0}\sin \varphi $, where $\varphi =0^{o}$ corresponds to the polarization
within the ab-plane.

\begin{figure}[th]
\centering
    \includegraphics[width=7cm,clip]{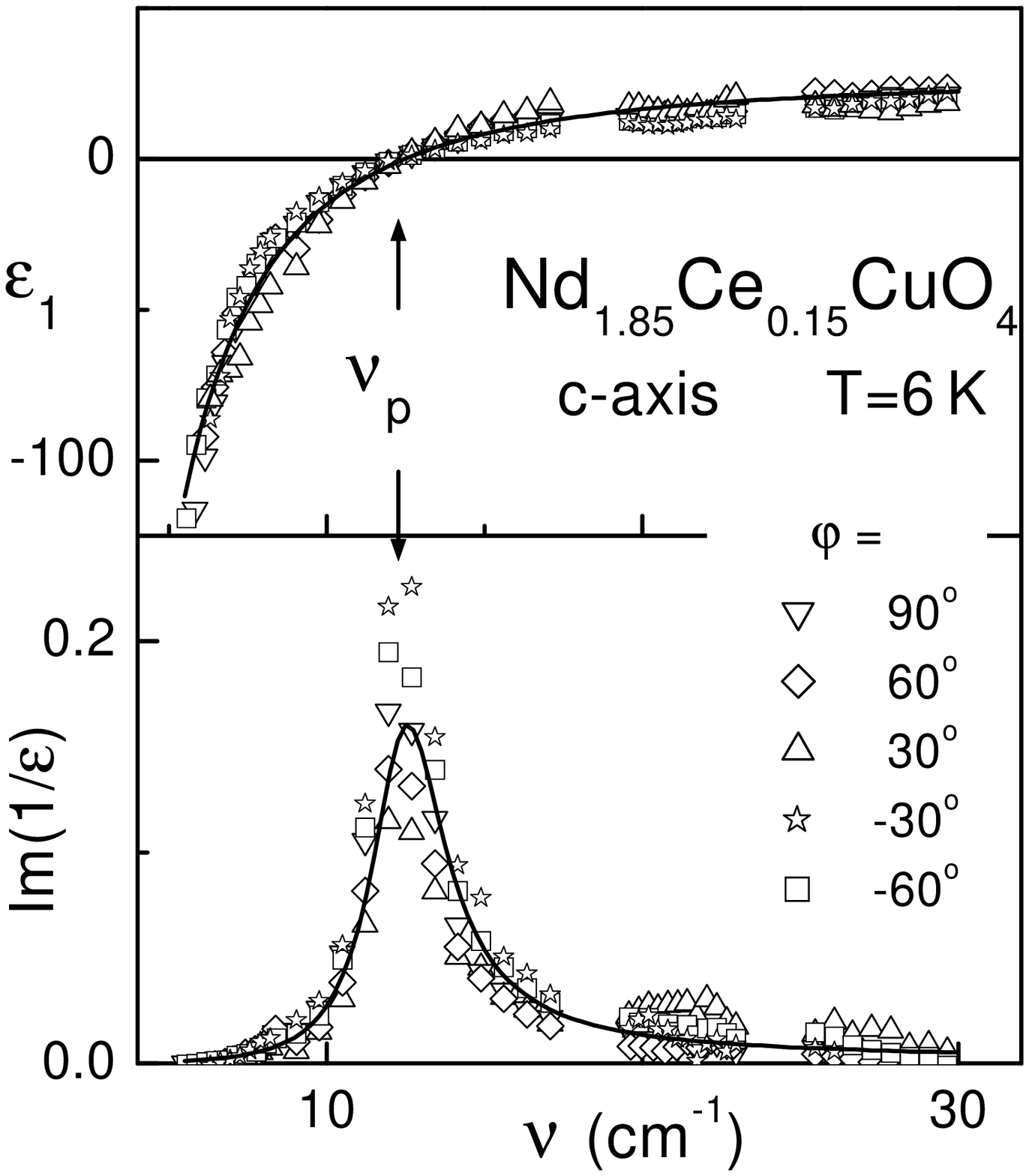}
    \vspace*{1mm}\caption{ Dielectric function of Nd$_{1.85}$Ce$_{0.15}$CuO$_{4-\delta}$ film
along the c-axis. Upper panel: $\varepsilon _1$. Lower panel: Imaginary part
of the loss function $Im(1/\varepsilon^{*})$. Lines are calculated within
the following assumptions: i) $\sigma _{1c}$ is frequency independent and
ii) only high frequency dielectric constant $\varepsilon _{\infty}$ and
superonducting condensate $-(\lambda _{c} c / \omega)^2$ contribute to $%
\varepsilon _1$ (see text).}
\label{figc}
\end{figure}

Using Eq. (1), it is possible to calculate the c-axis properties of Nd$%
_{1.85}$Ce$_{0.15}$CuO$_{4-\delta }$ film. The dielectric function along the
c-axis, $\varepsilon _{c}^{*}=i\sigma _{c}/\omega \varepsilon _{0}$,
obtained in this way, is presented in Fig. 3. Supporting the concept used,
the data for all polarizations $\varphi $ collapse into a single curve. It
should be noted, that the uncertainty in determination of $\alpha _{0}\,$%
represents the most relevant source of errors in the determination of $%
\varepsilon _{c}^{*}$. However, the variation of this angle influences only
the absolute values of the conductivity but leaves the overall frequency
dependence unchanged.

As demonstrated in Fig. 3, the high-frequency $(\nu >20$\thinspace cm$^{-1})$
c-axis dielectric constant of Nd$_{1.85}$Ce$_{0.15}$CuO$_{4-\delta }$ is
positive which can be ascribed to the phonon contribution. A superconducting
response becomes dominating for $\nu <15$\thinspace cm$^{-1}$ from which the
low-frequency penetration depth may be estimated as $\lambda _{c}(6K)=24\pm
8 $ $\mu $. From the zero crossing of the dielectric constant, the screened
plasma frequency can be determined as $\nu _{p}=12$\thinspace cm$^{-1}$.
This value is in excellent agreement with the powder data of Shibata and
Yamada\cite{shibata} and with recent results of Singley et al.\cite{basov}.

In conclusion, the anisotropic properties of a series of slightly tilted Nd$%
_{1.85}$Ce$_{0.15}$CuO$_{4-\delta }$ thin films were measures in the
submillimeter frequency range. Two new features were observed in the
low-temperature transmission spectra which showed different polarization
dependencies. The antiresonance at $\nu =12$\thinspace cm$^{-1}$ corresponds
to the c-axis plasma frequency of Nd$_{1.85}$Ce$_{0.15}$CuO$_{4-\delta }$.
The second resonance represents a mixed ab-plane/c-axis excitation. The
frequency of this resonance can be changed in a controllable way by rotating
the polarization of the incident radiation.

We acknowledge stimulating discussion with R. Hackl and W. N. Hardy.
This work was supported by BMBF (13N6917/0 - EKM)

\end{multicols}

\end{document}